\documentstyle[appb,epsfig]{article}

\newcommand{\barra}{\Big{/}}
\begin{document}

\begin{flushright}
%\begin{tabular}{r}
FTUV/98$-$59 \\ IFIC/98$-$60
%\end{tabular}
\end{flushright}

\vspace{0.5cm}

\begin{center}
{\Large {\bf INDIRECT CP VIOLATION IN THE $B_d$-SYSTEM\footnote{
Lecture presented at the Zeuthen Workshop on Elementary
Particle Physics ``Loops and Legs in Gauge Theories'', Rheinsberg,
April 19-24, 1998.}}}
\end{center}

\vspace{0.5cm}

\begin{center}
{\Large {J. Bernab\'eu$^a$ and M.C. Ba\~{n}uls $^b$
  }}
\end{center}

\vspace{0.5cm}

{\small {\it
$^a$ Departamento de F\'{\i}sica Te\'orica,  Universidad
de Valencia, 46100 Burjassot (Valencia), Spain.

$^b$   IFIC, Centro Mixto Universidad
de Valencia - CSIC, 46100 Burjassot (Valencia), Spain.

}}

\vspace{1.5cm}

\begin{abstract}
Recently a rephasing invariant definition of the CP-mixing parameter
for Indirect CP Violation has been introduced. This is made possible
by the explicit use of the CP operator into the  analysis. The problem
is then that of the determination  of the CP operator for a 
CP violating scenario. We discuss it and provide a definite solution.
\end{abstract}

\vspace{1cm}

\section{Introduction}

In 1964, Christenson, Cronin, Fitch and Turlay \cite{1} discovered
that the long-lived neutral kaon $K_L$ also decays to $\pi^+ \pi^-$
with a branching ratio of $\sim 2 \times 10^{-3}$. This discovery established
CP-Violation and the fact that $K_L$ is not identical to the CP-eigenstate
with CP-eigenvalue equal to $- 1$.

Similarly, the short-lived neutral kaon $K_S$ is not identical to the
CP-eigenstate with CP-eigenvalue equal to $+ 1$. CP Violation was confirmed
later by the decay $K_L \rightarrow \pi^0 \pi^0$ \cite{2} and by
the charge asymmetry \cite{3} in the $K_{l3}$ decays $K_L \rightarrow
\pi^{\pm} l^{\mp} \nu_l$. In particular, this semileptonic
asymmetry measures whether CP-violation is presented in the physical
eigenstates of the Meson Mass Matrix, referred to as Indirect CP 
Violation. The present value of the world average \cite{4} of the charge
asymmetry gives a CP-Violation in the Mixing $Re\,\varepsilon_K = (1.63 
\pm 0.06) \times 10^{-3}$.

For the non-leptonic $K_{S,L} \rightarrow 2 \pi$ decays, the experimentally
observable quantities are the ratios
\begin{equation}
\eta_\pm =  \frac{\langle \pi^+ \pi^- | K_L\rangle}{\langle \pi^+ \pi^- |
K_S \rangle}
\quad , \quad  \eta_{00} =  \frac{\langle \pi^0 \pi^0 | K_L\rangle}{\langle
\pi^0 \pi^0 | K_S \rangle}
\end{equation}
which can be rewritten in terms of CP-Violating indirect $\varepsilon$ and
direct $\varepsilon'$ parameters as
\begin{equation}
\eta_\pm \simeq \varepsilon + \varepsilon' \quad , \quad \eta_{00} \simeq
\varepsilon - 2 \varepsilon'
\end{equation}
when the $\Delta I = 1/2$ rule is used.

The ratio $\frac{\varepsilon'}{\varepsilon}$ can be determined by the
``method of ratio of ratios'' when comparing the $\pi^0 \pi^0$ and
$\pi^+ \pi^-$ decay channels
\begin{equation}
Re (\frac{\varepsilon'}{\varepsilon}) \simeq \frac{1}{6} \left\{
1 - \left| \frac{\eta_{00}}{\eta_{+-}} \right|^2 \right\}
\end{equation}

Present results at CERN \cite{5} and FermiLab \cite{6} give
conflicting results at the level of $10^{-3}$. New experiments with
better sensitivity at CERN, FermiLab and the dedicated $\Phi$-factory
at Frascati \cite{7} will push the precision to reach 
sensitivities better than $10^{-4}$. 

Recently the first direct observation of a difference in the decay
rates between particles and antiparticles has been accomplished by CP-LEAR
experiment \cite{8}. They make use of Flavour-Tag, either $K^0$ or
$\bar{K}^0$ at $t = 0$, and study their time evolution in the decays
to $2 \pi$. One concludes that, in the $K^0 - \bar{K}^0$ system, Indirect
CP Violation governed by  $\varepsilon_K$ plays the most prominent role.

The main question is whether the origin of CP Violation can be explained
within the Standard Model or it needs physics beyond the Standard Model.
In particular, is CP Violation to be described  by charged
current flavour mixing of quarks? We believe that the $B^0 - \bar{B}^0$
system is going to play a fundamental role in this respect and its
experimental study will be the task of all big facilities in the world:
B-factories, Cornell, HERA-B, B-TeV, LHC.

\section{Orthodoxy for Indirect CP Violation int he $B_d$-system.}

Flavour Number is not conserved by weak interactions, so in $2^{\mbox{nd}}$
order $B^0$ and $\bar{B}^0$ mix. The existence of decay channels leads
to a non-hermitian Mixing Matrix
\begin{equation}
H = M - \frac{i}{2} \Gamma
\end{equation}
in the $B^0, \bar{B}^0$ basis. The physical eigenstates of mass and
lifetime diagonalize $H$ as
\begin{equation}
\begin{array}{l}
| B_1 \rangle = p | B^0 \rangle + q | \bar{B}^0\rangle \\
| B_2 \rangle = p | B^0 \rangle - q | \bar{B}^0\rangle 
\end{array}
\end{equation}
with the amplitude
\begin{equation}
\frac{q}{p} = - \frac{2 M^*_{12} - i \Gamma_{12}^*}{\Delta m -
\frac{i}{2}  \Delta \Gamma} \quad ; \quad
M_{12} - \frac{i}{2} \Gamma_{12} \equiv \langle B^0 |H| \bar{B}^0 \rangle
\end{equation}
obtained in the Weisskopf-Wigner approximation.

\vspace{0.5cm}

The standard Model $\Delta B = 2$ transition is given by the Box
Diagram\footnote{This is my reference to the title of the Workshop}
\begin{figure}[htbp]
\begin{center}
\epsfig{file=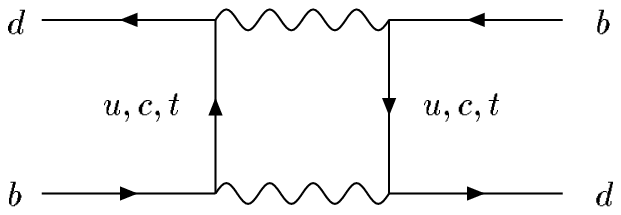}
\end{center}
\end{figure}

\noindent which generates matrix elements with $|\Gamma_{12}| << |M_{12}|$
and an almost alignment of the complex values of $\Gamma_{12} $ and $M_{12}$.
As a consequence, to a good approximation \cite{9}, the flavour mixing
amplitude $q/p$ is just a pure phase!

The parameter $q/p$ is phase-convention-dependent on the definition
of the CP-transformed states and thus its phase is  not, by itself, 
observable. The best prospects use then the strategy of the interplay
between Mixing and Decay \cite{10}.

The non-observability of the flavour mixing phase is made apparent in
the CP-violating rate asymmetry, from a \underline{flavour tag},
in the semileptonic decay $B^0 \rightarrow l \nu_l X$:
\begin{equation}
a_{SL} \equiv \frac{N (l^+ l^+) - N (l^- l^-)}{N (l^+ l^+) +
N (l^- l^-) } = \frac{| p/q|^2 - |q/p|^2}{|p/q|^2 + |q/p|^2}
\end{equation}

To generate $|q/p| \neq 1$, one would need both $\Delta \Gamma_B \neq 0$
and a misalignment such that $Im (M_{12}^* \Gamma_{12}) \neq 0$. Some
prospects could appear for physics beyond the Standard Model \cite{11}.

\section{Phase-convention-independent CP Mixing.}

The question is whether Indirect CP Violation and $|q/p| \neq 1$ are
equivalent. We propose to establish the concept of Indirect CP
Violation by means of CP Mixing in the physical states. We define the
$\varepsilon$-parameter \cite{12} as 
\begin{equation}
\left.
\begin{array}{l}
|B_1 \rangle =
\frac{1}{\sqrt{1 + |\varepsilon|^2}} (|B_+ \rangle + \varepsilon |B_- \rangle)\\[2ex]
|B_2 \rangle =
\frac{1}{\sqrt{1 + |\varepsilon|^2}} (|B_- \rangle + \varepsilon |B_+ \rangle)\\
\end{array}  \right\} \quad
|B_\pm \rangle \equiv \frac{1}{\sqrt{2}} (I \pm CP) |B^0 \rangle
\end{equation}
where $|B_\pm \rangle$ are the CP eigenstates. For a given CP-opertor, this
$\varepsilon$ is phase-convention-independent. This is seen explicity by
the relation 
\begin{equation}
\frac{1 - \varepsilon}{1 + \varepsilon} = \frac{q}{p} CP_{12}
\end{equation}
and the result (6): The phase convention for $B^0, \bar{B}^0$ states
is irrelevant. $\varepsilon$ involves, on the other hand, the three operators
$M, \Gamma$ and $CP$. the matrix element $CP_{12} \equiv \langle B^0 |CP| 
\bar{B}^0 \rangle$ plays the role of a reference phase, given by the flavour
mixing amplitude in the $CP$ conserving limit
\begin{equation}
\left(\frac{q}{p}\right)_{CP} = - CP_{12}^*
\end{equation}

With three directions in the complex plane, those of $M_{12}, \Gamma_{12},
CP_{12}$, we have two \underline{relative} phases that can become
observable: one is well known
\begin{equation}
\frac{2 Re (\varepsilon)}{ 1 + |\varepsilon|^2} = \frac{1 - |q/p|^2}{1 +
|q/p|^2}
\end{equation}
which involves the relative phase between $M_{12} $ and $\Gamma_{12}$.
For the $B_d$-system, with $\Delta \Gamma \sim 0$, one has $Re
(\varepsilon)\sim 0$. In this limit,
\begin{equation}
\frac{Im (\varepsilon)}{1 + |\varepsilon|^2} \simeq \frac{Im
(M_{12}^* CP_{12})}{\Delta m}
\end{equation}
and we observe that the (second) relative phase between $M_{12}$
and $CP_{12}$ gives $Im (\varepsilon)$. Is this a quantum-mechanical
observable?

In Ref. \cite{12} we have discussed and interference experiment between
the CP-eigenstates $| B_{\pm} \rangle$ obtained from the time evolution of a
CP-Tag. The corresponding CP asymmetries to common leptonic final 
states $l^+$ and $l^-$ are able to separate out $Re (\varepsilon)$ and
$Im (\varepsilon)$. If $B_+ (t)$ denotes the time-evolved state
from a CP eigenstate $B_+$ prepared at $t = 0$, one has 
\begin{equation}
\begin{array}{l}
A_+^{CP} (t) =
\frac{\displaystyle{\Gamma[B_+ (t) \rightarrow l^+] - \Gamma [B_+
(t) \rightarrow l^-]}}{\displaystyle{\Gamma [B_+ (t) \rightarrow l^+] +
\Gamma [B_+ (t) \rightarrow l^-]}}\\[2ex]
= \frac{\displaystyle{2 Re (\varepsilon)}}{
\displaystyle{1 + |\varepsilon|^2}} [1 - e^{\frac{\Delta \Gamma}{2} t}
\cos (\Delta m\;t)] - \frac{\displaystyle{2 Im (\varepsilon)}}{
\displaystyle{1 + |\varepsilon|^2}}
e^{\frac{\Delta \Gamma}{2} t} \sin (\Delta m\; t)
\end{array}
\end{equation}

We conclude that $Im (\varepsilon)$ is a physical quantity and observable
iff the CP-transformation is well defined. Is the CP-operator determined?
In the case of invariant theory, a symmetry transformation is well defined.
Otherrwise the corresponding symmetry operator is undetermined.

\section{The CP-conserving direction.}

It is possible to have a well defined CP operator even for a non invariant
theory. This is the situation when the structure of the Lagrangian
allows the separation of a CP conserving part which includes flavour
mixing, from a different interaction responsible for the CP non-invariance.
In this type of models, as in superweak interaction \cite{13}, the
invariant part of the Lagrangian determines the action  of the symmetry
operation on the fields.

The most interesting case is that of the Standard Model: a theory where
flavour mixing and CP violation cannont be separated in that way.
Therefore, there is no phase choice for the CP trasformed fields which 
leaves the Lagrangian invariant. Different choices of phases, and thus
of CP operator, will yield different observables: our $\varepsilon$-parameter
would not be unique. We are going to show, however, that the use of the
quark mixing hierarchy (empirically well established) leads to a unique
separation  of the weak Lagrangian into a CP-conserving and a CP-violating
part.

The CP operation is defined by the invariance of strong and 
electromagnetic interactions. When the mass matrices for the up $M$ and
down $M'$ sectors are considered, the corresponding electroweak quark
fields have CP transformations which include unitary matrices
$\Phi$ and $\Phi \, '$, respectively, in family space. The invariance condition
on the Lagrangian determines \cite{14} $\Phi, \Phi\, '$ up to diagonal
unitary phases  $e^{2 i \theta}, e^{2i \theta'}$ in terms of the diagonalizing
matrices $U, U'$ for the quark fields. These diagonal phases are the
(arbitrary) CP-phases of the physical quark fields. If the charged 
current Lagrangian was absent, there would be no cross-talk 
between up and down quarks: the arbitrariness would have no physical
effect and we could consider the CP operator to be determined.

On the contrary, the existence of a charged current Lagrangian
induces both

- Flavour Mixing through $V = U {U'}^\dagger$

- CP Violation through $B = \Phi {\Phi\,'}^\dagger$

The arbitrariness of CP-phases in now relevant to induce different
 $B$-matrix. Is it possible to choose $\theta,  \theta'$ such
that $B = I$? If the answer is positive, we have a theory with Flavour
Mixing $V \neq I$, but CP Invariance $B = I$: a CP conserving Standard
Model. CP Violation would have to be understood from a Superweak-type
Model. The necessary and sufficient conditions for CP invariance
to be satisfield \cite{15} by the mass matrices $M, M'$ and the
corresponding phase fixing were discussed \cite{16} some time ago.

If there is no CP-phase choice for $\theta, \theta'$ to get $B = I$, the
theory is CP-violating. Can the theory still filter a well defined CP
operator, at least in a perturbative sense? We know that, in the
K-system, the  CP symmetry is only slightly violated and its size
\cite{1} is of the order $ O (10^{-3})$. This is understood in the
Standard Model as a consequence of the need to involve the three
families to generate CP violation. Thus its effective coupling 
contains higher powers of the quark mixing $\lambda$ than that
of the CP conserving flavour mixing $K^0 \bar{K}^0$. This justifies
the idea to look for a ``natural'' CP definition in the Standard Model
based on the empirically known quark mixing hierarchy.

Take one of the sides (fixed $k$) of the $(bd)$ unitarity triangle.
It can be decomposed  in the (complex) plane into CP conserving and CP
violating parts as
\begin{equation}
\begin{array}{l}
V_{kb}^* V_{kd} (CP) = e^{i (\theta_b - \theta_d)} Re
(e^{-i \theta_b}
V_{kb}^* V_{kd} e^{i \theta_d}) \\[2ex]
V_{kb}^* V_{kd} ({{CP}\kern-13pt\raise+0.7ex\hbox{\barra}}\;\;)
= i e^{i (\theta_b - \theta_d)} Im (e^{-i \theta_b}
V_{kb}^* V_{kd} e^{i \theta_d}) 
\end{array}
\end{equation}

Eq (14) tells that $e^{i (\theta_b - \theta_d)}$ defines the
CP conserving direction associated to the ($bd)$ triangle. It depends
on the choice of CP-phases. However, the three CP conserving directions
of the three ``down'' triangles are not independent, due to the cyclic
relation

\begin{equation}
e^{i (\theta_b - \theta_d)} = e^{i (\theta_b - \theta_s)} e^{i (\theta_s -
\theta_d)}
\end{equation}

These CP conserving directions are attached to the triangles, so they
would rotate with them under quark rephasing. They are not physical
by themselves, but the relative phases between triangle sides and them
are rephasing invariant.

According to the experimentally known hierarchy in the quark mixing,
the magnitude of $V$ matrix elements can be written in terms of a 
perturbative parameter $\lambda$. We can estimate the relative
size of every side in the three triangles of the down sector. To order
$\lambda^3$, the two triangles ($bs$) and $(sd)$ collapse to
a line each, thus giving  CP conservation and a natural choice for the
attached CP direction: the CP invariance requirement on the effective
hamiltonian fixes the corresponding CP-phases of these sectors. Due
to (15) the CP conserving direction for the ($bd)$ system is already
fixed. This is particularly attractive because the ($bd$) system
keeps a CP violating triangle to order $\lambda^3$. One obtains \cite{14}

\begin{equation}
e^{i (\theta_b - \theta_d)} = \left. \frac{V_{cd} V^*_{cb}}{|V_{cd} 
V^*_{cb}|}\,
\right|_{O (\lambda^3)}
\end{equation}

Thus the CP conserving direction matches one of the sides of the ($bd)$
triangle to $O (\lambda^3)$.

This result controls the value of $Im (\varepsilon)$ for the $B_d$-system
as an OBSERVABLE even in the Standard Model.

To order $\lambda^3$, we have

\begin{equation}
Im (M^*_{12} CP_{12})  \propto Im (V^*_{td} V_{tb} V_{cd} V^*_{cb})
\end{equation}

Eq. (17) proves definitively the phase-convention-independence
of our analysis.

\section{Conclusions.}

These are summarized in the following two points:

1) There exists a rephasing invariant measure of CP Mixing given by our
$\varepsilon$-parameter. It is independent of the rephasing of $B^0 -
\bar{B}^0$ states and of quark fields, i.e., independent of a specific
parametrization of the Mixing Matrix $V (CKM)$.

2) The $\varepsilon$-parameter is unique iff the CP operator is well
defined. The use of the quark mixing hierachy leads to determine CP
to $O (\lambda^3)$.

As a final comment, the definite CP conserving direction found
in Eq. (16) implies that decays of $B_d$ to final CP eigenstates
dominated by the amplitude $V_{cd} V_{cb}^*$ constitute excellent CP
Tags.

\vspace{3cm}

\noindent
\underline{Acknowledgements}

J.B. would like to acknowledge the splendid scientific atmosphere
of the Workshop. Special thanks go to Tord Riemann for his dedication. 
M.C.B. is indebted to the Spanish Ministry of Education and Culture
for her fellowship. This work was supported by CICYT, Spain, under
Grant AEN-96/1718.


\newpage

\begin{thebibliography}{99}
\bibitem{1} J.H. Christenson, J.W. Cronin, V.L. Fitch and R. Turlay,
Phys. Rev. Lett. 13 (1964) 138.
\bibitem{2} J. W. Gaillard {\em et al.}, Phys. Rev. Lett. 18 (1967) 20;
J.W. Cronin {\em et al.}, Phys. Rev. Lett. 18 (1967) 25.
\bibitem{3} S. Bennet {\em et al.}, Phys. Rev. Lett. 19 (1967) 993; D. Dorfan
{\em et al.}, Phys. Rev. Lett. 19 (1967) 987.
\bibitem{4} Particle Data Book, Review of Particle Properties, Phys.
Rev. D54 (1996) 1.
\bibitem{5} G.D. Barr {\em et al.}, NA31, Phys. Lett. B317 (1993) 233.
\bibitem{6} L.K. Gibbons {\em et al.}, E731, Phys. Rev. Lett. 70 (1993) 1203.
\bibitem{7} See the Lecture by L. Pancheri in these Proceedings.
\bibitem{8} R. Adler {\em et al.}, CPLEAR, Phys. Lett. B363 (1995) 243.
\bibitem{9} V. Khoze, M. Shifman, N. Uraltsev and M. Voloshin, Yad.
Fiz. 46 (1987) 181.
\bibitem{10} I.I. Bigi {\em et al.} in ``CP Violation'', p. 175., ed. 
C. Jarlskog (World Scientific, Singapore, 1989).
\bibitem{11} G.C. Branco, T. Morozumi, P.A. Parada and M.N. Rebelo,
Phys. Rev. D48 (1993) 1167; G. Barenboim, J. Bernab\'eu and M. Raidal, 
Nucl. Phys. B511 (1998) 577.
\bibitem{12} M.C. Ba\~{n}uls and J. Bernab\'eu, Phys. Lett B423 (1998) 151.
\bibitem{13} L. Wolfenstein, Phys. Rev. Lett. 13 (1964)562.
\bibitem{14} M.C. Ba\~{n}uls and J. Bernab\'eu, FTUV 98/37 (1998).
\bibitem{15} C. Jarlskog, Phys. Rev. Lett. 55 (1985) 1039; Z. Phys.
C29 (1985) 491.
\bibitem{16} J. Bernab\'eu, G. Branco and M. Gronau, Phys. Lett. B169
(1986) 243.
\end{thebibliography}
\end{document}